\documentstyle[aps,epsf]{revtex}

\def \beq {\begin{equation}}
\def \eeq {\end{equation}}
\def \bes {\begin{eqnarray}}
\def \ees {\end{eqnarray}}
\def\ni{\noindent}
\def\nn{\nonumber}

\begin{document}
\title{
Surface impedance and the Casimir force
}

\author{
V.~B.~Bezerra,
G.~L.~Klimchitskaya\footnote{On leave from
North-West Polytechnical University,
St.Petersburg, Russia.\\
E-mail: galina@fisica.ufpb.br}
and C.~Romero
}
\address
{
 Departamento de F\'{\i}sica, Universidade Federal da Para\'{\i}ba,\\
Caixa Postal 5008, CEP 58059-970,
Jo\~{a}o Pessoa, Pb-Brazil
}

\maketitle

\begin{abstract}
The impedance boundary condition is used to calculate the Casimir
force in configurations of two parallel plates and a shpere (spherical 
lens) above a plate at both zero and nonzero temperature. The impedance
approach allows one to find the Casimir force between the realistic
test bodies regardless of the electromagnetic fluctuations
inside the media. Although this approach is an approximate one, it 
has wider areas of application than the Lifshitz theory of
the Casimir force. The general formulas of the impedance approach
to the theory of the Casimir force are given and the formal
substitution is found for connecting it with the Lifshitz formula.
The range of micrometer separations between the test bodies 
which is interesting from the experimental point of view 
is investigated in detail.
It is shown that at zero temperature the results obtained on the basis of
the surface impedance method are in agreement with those obtained 
in framework 
of the Lifshitz theory within a fraction of a percent. The 
temperature correction to the Casimir force 
from the impedance method
coincides with that from the Lifshitz
theory up to four significant figures. The case of
millimeter separations which corresponds to the
normal skin effect 
is also considered. At zero temperature the obtained results have good 
agreement with the Lifshitz theory. At nonzero temperature the impedance 
approach is not subject to the interpretation problems peculiar to the
zero-frequency term of the Lifshitz formula in dissipative media.
\end{abstract}




\section{Introduction}

Currently the Casimir effect \cite{1} has been attracted much attention
owing to its promising applications in both fundamental physics and
nanotechnology (see the recent review \cite{2}). In its simplest form 
the Casimir effect consists in the appearance of a force  between
two parallel uncharged metallic plates separated by a distance $a$.
This is a purely quantum effect caused by the alteration of the zero-point 
oscillations of the electromagnetic field due to the presence
of the plates. Experimentally, very precise measurements of the Casimir
force have been performed recently 
by several authors \cite{3,4,5,6,7,8,9}.
Theoretically, different configurations have been 
considered and the role of more realistic 
conditions, such as nonzero temperature \cite{10,11,12}, finite
conductivity of the boundary metal \cite{13,14,15,16} and surface
roughness \cite{17,18,19}, carefully examined.

Fundamental applications of the Casimir effect belong to the domains of
the Kaluza-Klein supergravity, quantum chromodynamics, atomic physics and
condensed matter \cite{2,20,21}. In the last few years it has 
been used to obtain
stronger constraints on the constants of hypothetical long-range
interactions \cite{22,23,24}. Concerning nanotechnology 
the first microelectromechanical system actuated by the Casimir effect
should be mentioned \cite{9}.
The nontrivial boundary dependence of the Casimir force (see, e.g.,
\cite{5,25,26}) opens up new possibilities of using the Casimir effect in
technology.

All the above-listed applications require precise methods of calculation of
the Casimir force which take account of different influential factors and
could be applied to various configurations. The most fundamental basis
for the calculation of the Casimir force for realistic media is
suggested by Lifshitz theory \cite{27,28}. According to this theory
the vacuum oscillations are modeled by a randomly fluctuating
electromagnetic field propagating inside a dielectric material described
by a frequency-dependent dielectric permittivity. Lifshitz theory gives the
possibility of calculating the Casimir force taking into account 
both nonzero 
temperature and finite conductivity of the boundary metal 
\cite{10,11,12,13,14,15}. It faces some problems, however, 
concerning the interpretation 
of the zero-frequency contribution to the Casimir force between metals at
nonzero temperature \cite{12}.

The other method that permits to calculate the Casimir force between the
realistic media is based on the use of impedance boundary conditions.
This method allows one to find the Casimir force between realistic media
similarly to the way it was found in Quantum Field Theory, i.e. by imposing
some boundary conditions on the bounding surfaces and not considering
the regions occupied by the media. The advantages of the impedance
method are the following. It gives the possibility to
carry out all calculations more simply. It also permits to obtain results
in the cases Lifshitz theory has problems due to indefinite
value of the zero-frequency term \cite{12} or when the characteristic 
frequencies correspond to the anomalous skin effect, or in the case of 
low-temperature superconductors where the dielectric permittivity is not
a well-defined quantity \cite{20}.

Surface impedance was first applied in Ref.~\cite{29} to calculate the
Casimir force acting between two metallic plates made of real metal at
zero temperature (a similar in spirit approach was used in \cite{30}
where the reflection coefficients of electromagnetic waves from metal
plates were approximatelly expressed in terms of the impedance function).
Recent trends increasingly make use of impedance boundary conditions in
computer simulation studies of the scattering of light from metal films
\cite{31,32,33}. The elimination of the electromagnetic field inside the
scattering medium by the use of an impedance boundary condition leads
to a significant reduction of computation time with no notable loss of
accuracy \cite{31,32,33}. Furthermore, the use of impedance makes it
possible to consider curved or rough surfaces.

In this paper we first apply impedance boundary conditions to calculate
the Casimir force between real metals at nonzero temperature. 
Both the configurations of two plates and a sphere (spherical lens) above
a plate are considered. The separations between the test bodies are
examined from one tenth of a micrometer to around a hundred micrometers 
which cover the visible spectrum and infrared optics. The smallest
separations of this interval correspond to the area of interest of 
modern experiments \cite{3,4,5,6,7,8,9} on measuring the Casimir force.
At zero temperature the impedance approach agrees with the Lifshitz 
formula up to a fraction of a percent.
It is shown that the results for the temperature correction to the
Casimir force obtained via the surface impedance are in agreement with
those obtained in frames of Lifshitz theory up to four significant
figures.
The region of the normal skin effect is also examined, which corresponds
to the space separations between test bodies larger than one
millimeter. This case is interesting from a methodological point of view
because in the presence of dissipation at nonzero temperature the
zero-frequency term of the Lifshitz formula has an indefinite
character \cite{12}. The impedance approach was demonstrated to have
no indefinite contributions. It agrees with the Lifshitz formula at
zero temperature in the region of the normal skin effect and leads 
to quite reasonable results at nonzero temperature.

The paper is organized as follows. In Sec.II the main points of 
the impedance 
approach to the theory of the Casimir force acting between real media
are outlined. Sec.III contains calculations of the Casimir force at zero 
temperature in the separation range corresponding to visible light and 
infrared optics, and the comparison of the obtained results with those
found by considering Lifshitz theory. In Sec.IV the Casimir force at
nonzero temperature is computed in the region of visible spectrum and
infrared optics. The excellent agreement of the obtained 
temperature correction
with the results based on Lifshitz theory is shown. Sec.V is
devoted to numerical and analytical calculations of the Casimir force in
the region of the normal skin effect. Both cases of zero and nonzero 
temperature are considered. In Sec.VI the reader will find conclusion
and discussion.

\section{Impedance approach to the theory of the Casimir effect}

It is common knowledge \cite{34} that the penetration of an electromagnetic
field into a real metal can be effectively described by imposing an
impedance condition on the boundary surface $\Gamma$
\beq
{\mbox{\boldmath$E$}}_t{\vert}_{\Gamma}=Z(\omega)
\left({\mbox{\boldmath$H$}}_t\times{\mbox{\boldmath$n$}}
\right){\vert}_{\Gamma}.
\label{1}
\eeq
\ni
Here $Z(\omega)$ is the impedance, {\boldmath$n$} is the internal
normal to the boundary surface and the lower index $t$ denotes the
tangential component of the electric and magnetic fields.
If a non-magnetic medium can be characterized by some definite
dielectric permittivity $\varepsilon(\omega)$, then the impedance is
expressed in terms of it as
\beq
Z(\omega)=1/\sqrt{\varepsilon(\omega)}.
\label{2}
\eeq
\ni
It is known that Re$Z$ is proportional to the time-averaged energy
flux through the boundary surface. For an ideal metal
($\varepsilon=\infty$ at all frequencies) it follows that $Z(\omega)=0$
and we return from (\ref{1}) to Dirichlet boundary condition
that is most often used in the field-theoretical approach to the
theory of the Casimir effect.

To find the eigenfrequencies of the problem one should solve Maxwell
equations with the condition (\ref{1}) on the boundary surfaces.
The Casimir energy can be obtained as 
half the sum of these eigenfrequencies, 
which is most simply calculated by the use of the argument theorem.
By this means the result is expressed in terms of the impedance function
at the imaginary frequencies. The subtraction of infinities is carried 
out by requiring that for the infinitely remote boundary surfaces the
physical vacuum energy must be zero. The above-mentioned calculation
was performed in \cite{29} (see also \cite{20}) 
at zero temperature for the case of two
semispaces modelling two parallel plates separated by an empty gap
$-a/2\leq z\leq a/2$. The result for the energy per unit area is
\beq
E_{pp}^{(0)}(a)=\frac{\hbar}{(2\pi)^2}
\int_{0}^{\infty} d\zeta
\int_{0}^{\infty} Q\,dQ
\ln\left(D^{\|}D^{\bot}\right),
\label{3}
\eeq
\ni
where {\boldmath$Q$} is a two-dimensional wave vector in the plane of 
plates, $Q=|\mbox{\boldmath$Q$}|$, the frequency on the imaginary axis is 
$\omega=i\zeta$, and the upper index $(0)$ refers to the case
of zero temperature. The explicit form of functions $D$ for parallel and
perpendicular polarizations, respectively, is given by
\beq
D^{\|}=D^{(0)}\left[1+\frac{4\zeta RZ(i\zeta)}{(R+\zeta Z(i\zeta))^2}
\frac{1}{e^{2Ra}-1}\right],
\qquad
D^{\bot}=D^{(0)}\left[1+\frac{4\zeta RZ(i\zeta)}{(\zeta+R Z(i\zeta))^2}
\frac{1}{e^{2Ra}-1}\right].
\label{4}
\eeq
\ni
Here the following notations are introduced
\beq
D^{(0)}\equiv 1-e^{-2Ra}, \qquad R\equiv\sqrt{\frac{\zeta^2}{c^2}+Q^2}.
\label{5}
\eeq
\ni
Note that for ideal metal $Z(i\zeta)=0$, $D^{\|}=D^{\bot}=D^{(0)}$
and Eq.(\ref{3}) immediately leads to
\beq
E_{pp}^{(0,0)}(a)=
\frac{\hbar}{2\pi^2}
\int_{0}^{\infty} d\zeta
\int_{0}^{\infty} Q\,dQ
\ln\left(1-e^{-2Ra}\right)
=-\frac{\pi^2\hbar c}{720a^3},
\label{6}
\eeq
\ni
which is indeed correct for the ideal metal \cite{2,20,21} (the second upper
index denotes the case of ideal metal).

For further computation it is convenient to introduce the dimensionless
variables
\beq
y=2Ra, \qquad \xi=\frac{2a\zeta}{c}.
\label{7}
\eeq
\ni
In terms of these variables Eqs.(\ref{3}) and (\ref{4}) take the form
\beq
E_{pp}^{(0)}(a)=\frac{\hbar c}{32\pi^2a^3}
\int_{0}^{\infty} d\xi
\int_{\xi}^{\infty} y\,dy
\left\{\vphantom{\ln\left[1+\frac{X^{\|}(y,\xi)}{e^y-1}\right]}
2\ln\left(1-e^{-y}\right)\right.
+\left.
\ln\left[1+\frac{X^{\|}(y,\xi)}{e^y-1}\right]+
\ln\left[1+\frac{X^{\bot}(y,\xi)}{e^y-1}\right]\right\},
\label{8}
\eeq
\ni
where the quantities $X^{\|,\bot}(y,\xi)$ are given by
\beq
X^{\|}(y,\xi)=\frac{4\xi yZ}{(y+\xi Z)^2},\quad
X^{\bot}(y,\xi)=\frac{4\xi yZ}{(\xi+y Z)^2},
\qquad
Z\equiv Z\left(i\frac{c\xi}{2a}\right).
\label{9}
\eeq
\ni
Note that the first contribution in the right-hand side of Eq.(\ref{8})
describes the case of an ideal metal (see Eq.(\ref{6})). 
This integral can be but will not be carried out explicitly 
since in the case of nonzero temperature, considered below,  
the representation of (\ref{8}) it terms of the integrals is more appropriate.

The Casimir force per unit area of the plates is obtained from
Eq.(\ref{8}) as
\beq
F_{pp}^{(0)}(a)=-\frac{\partial E_{pp}^{(0)}(a)}{\partial a}=
-\frac{\hbar c}{32\pi^2a^4}
\int_{0}^{\infty} d\xi
\int_{\xi}^{\infty} y^2\,dy
\left[\frac{1-X^{\|}(y,\xi)}{e^y-1+X^{\|}(y,\xi)}+
\frac{1-X^{\bot}(y,\xi)}{e^y-1+X^{\bot}(y,\xi)}\right].
\label{10}
\eeq
\ni
It is notable that if we substitute the quantities $X^{\|,\bot}$, as
defined in Eq.(\ref{9}), for
\beq
X_L^{\|}(y,\xi)=\frac{4yZ\sqrt{\xi^2+(y^2-\xi^2)Z^2}}{(y+
Z\sqrt{\xi^2+(y^2-\xi^2)Z^2})^2},
\qquad
X_L^{\bot}(y,\xi)=\frac{4yZ\sqrt{\xi^2+(y^2-\xi^2)Z^2}}{(y Z+
\sqrt{\xi^2+(y^2-\xi^2)Z^2})^2},
\label{11}
\eeq
\ni           
and take Eq.(\ref{2}) into account, then Eqs.(\ref{8}), (\ref{10})
 would coincide with the Lifshitz result for the Casimir energy
density and force.  What this means is that the impedance approach
($X^{\|,\bot}$ are defined by Eq.(\ref{9})) is obtained from the
Lifshitz formula if we put $y^2-\xi^2=0$ in Eq.(\ref{11}).
Taking Eqs.(\ref{5}) and (\ref{7}) into account one
can conclude that the impedance approximation to Lifshitz theory
corresponds to the dominant contribution of the photons with a small
wave vector in the plane of plates.

The results obtained for  the configuration of two parallel plates
can be easily adapted for the configuration of a sphere (spherical lens)
above a plate. For this purpose the proximity force theorem \cite{36}
can be used.  According to  this theorem the force acting between a sphere
and a plate is expressed via the energy density between 
the two parallel plates as
\bes
&&
F_{ps}^{(0)}(a)=2\pi RE_{pp}^{(0)}(a)=
\frac{\hbar cR}{16\pi a^3}
\int_{0}^{\infty} d\xi
\int_{\xi}^{\infty} y\,dy
\label{12}\\
&&
\phantom{aaaa}\times
\left\{2\ln\left(1-e^{-y}\right)+
\ln\left[1+\frac{X^{\|}(y,\xi)}{e^y-1}\right]+
\ln\left[1+\frac{X^{\bot}(y,\xi)}{e^y-1}\right]\right\}.
\nn
\ees
\ni   
For the configurations with small deviations from plane parallel geometry
the exactness of the proximity force theorem and the other
approximative methods is very high.
Thus for the two plates inclined at a small angle one to another 
the results obtained by the use of the additive summation 
method  differ from the exact ones
by less than $10^{-2}$\% \cite{37}. For the configuration of a sphere
above a plate the error introduced by the proximity  force theorem
is of order $a/R$ \cite{38}. To illustrate, that is of the order of 
0.1\% for the experimental parameters of Refs.\cite{4,5,6,7}.

All the above results were formulated for the case  of zero temperature.
It is not difficult, however, to modify  them for the case of
nonzero temperature $T$. In order to do so one must change the
integration with respect to continuous frequency in Eq.(\ref{6}), as
well as in Eqs.(\ref{8}), (\ref{10}), (\ref{12}) for the summation over
the Matsubara frequencies
\beq
\zeta_l=2\pi\frac{k_BT}{\hbar}l,
\label{13}
\eeq
\ni
where $l=0,1,2,\ldots\,$, and $k_B$ is the Boltzmann constant. In accordance
with Eq.(\ref{7}) the dimensionless Matsubara frequencies are given by
\beq
\xi_l=2\pi\frac{T}{T_{eff}}l,\qquad k_BT_{eff}=\frac{\hbar c}{2a},
\label{14}
\eeq
\ni
$T_{eff}$ being the effective temperature. As a result, the  Casimir
energy density between two plates at temperature $T$ in the impedance
approach takes the form
\beq
E_{pp}^{(T)}(a)=
E_{pp}^{(T,0)}(a)+
\frac{k_BT}{8\pi a^2}
\sum\limits_{l=0}^{\infty}{\vphantom{\sum}}^{\prime}
\int_{\xi_l}^{\infty} y\,dy
\left\{
\ln\left[1+\frac{X^{\|}(y,\xi_l)}{e^y-1}\right]\right.
+\left.
\ln\left[1+\frac{X^{\bot}(y,\xi_l)}{e^y-1}\right]\right\}.
\label{15}
\eeq
\ni         
Here the prime near a summation sign  means that the multiple
1/2 is added to the term with $l=0$. The quantity $E_{pp}^{(T,0)}$
is the Casimir energy density at temperature $T$ between the plates
made of ideal metal. The explicit expression for it is well known
\cite{2,39,40}:   
\bes
&&
E_{pp}^{(T,0)}(a)=
\frac{k_BT}{4\pi a^2}
\sum\limits_{l=0}^{\infty}{\vphantom{\sum}}^{\prime}
\int_{\xi_l}^{\infty} y\,dy
\ln\left(1-e^{-y}\right)
\label{16}\\
&&=
E_{pp}^{(0,0)}(a)\left\{1+
\frac{45}{\pi^3}
\sum\limits_{n=1}^{\infty}
\left[\left(\frac{T}{T_{eff}}\right)^3\frac{1}{n^3}
\coth\left(\pi n\frac{T_{eff}}{T}\right)\right.\right.
\nn \\
&&\phantom{aaaa}
\left.\left.
+\pi\left(\frac{T}{T_{eff}}\right)^2\frac{1}{n^2}
{\sinh}^{-2}\left(\pi n\frac{T_{eff}}{T}\right)\right]
-\left(\frac{T}{T_{eff}}\right)^4
\vphantom{\sum\limits_{n=1}^{\infty}}\right\},
\nn
\ees
where the energy density between two plates made of ideal
metal at zero temperature, $E_{pp}^{(0,0)}$, was defined in
Eq.(\ref{6}). The force acting between a sphere (spherical lens)
and a plate at nonzero temperature is obtained from Eqs.(\ref{15}),
(\ref{16}) by multiplying the right-hand sides by $2\pi R$
(compair with Eq.(\ref{12})).

In the impedance approach the force acting between two plates
made of real metal at temperature $T$ is obtained from Eq.(\ref{10})
by the substitution of (\ref{14}) and results in 
\beq
F_{pp}^{(T)}(a)=
-\frac{k_BT}{8\pi a^3}
\sum\limits_{l=0}^{\infty}{\vphantom{\sum}}^{\prime}
\int_{\xi_l}^{\infty} y^2\,dy
\left[\frac{1-X^{\|}(y,\xi_l)}{e^y-1+X^{\|}(y,\xi_l)}
+\frac{1-X^{\bot}(y,\xi_l)}{e^y-1+X^{\bot}(y,\xi_l)}\right].
\label{17}
\eeq

We conclude this section by stressing that  the above method
employed 
to get the Casimir force at nonzero temperature by changing
integration for summation according to Eqs.(\ref{13}), (\ref{14})
is valid under certain assumptions only. Specifically, the function
under the integrals in Eqs.(\ref{8}), (\ref{10}) must be 
well-defined at all points $(\xi_l,y)$, including the point (0,0).
The problems arising when this is not the case are detailly
discussed in Refs.\cite{12,41}.

\section{Casimir effect in the region of visible light and
infrared optics at zero temperature}

Here we apply the impedance approach to calculate the Casimir
energy density and force when the characteristic
frequencies giving the major contribution to the effect fall
into the region of visible light and infrared optics. In fact,
we will consider
the space separations between the test bodies from one tenth
of a micrometer to around a hundred micrometers.
For the characteristic frequencies under consideration, as the
field penetrates into the metal, it decays exponentially.
The surface impedance function is pure imaginary and given
by \cite{42}
\beq
Z(\omega)=-\frac{i\omega}{c}\delta(\omega)=
-\frac{i\omega}{\sqrt{\left(\frac{\varepsilon_F}{\hbar}\right)^2-
\omega^2}},
\label{18}
\eeq
\ni
where $\delta(\omega)$ is the depth of a skin layer and
$\varepsilon_F$ is the Fermi energy. It is suggested that
$\omega<\varepsilon_F/\hbar$. The frequency $\varepsilon_F/\hbar$
usually belongs to the shortwave optical or near-ultraviolet
parts of spectra. We consider below the case of a spherical Fermi 
surface when $\varepsilon_F=\hbar\omega_p$, $\omega_p$ being the
effective plasma frequency. Performing the rotation to the imaginary
frequency axis and using the dimensionless frequency, introduced
in Eq.(\ref{7}), one obtains
\beq
Z\equiv Z\left(\frac{ic\xi}{2a}\right)=
\frac{\xi}{\sqrt{{\tilde{\omega}}_p^2+\xi^2}},
\label{19}
\eeq
\ni
where ${\tilde{\omega}}_p=2a\omega_p/c$.

If the relevant frequencies obey the inequality 
$\omega\ll\varepsilon_F/\hbar$ 
($\xi\ll{\tilde{\omega}}_p$)
the impedance function takes a simpler form \cite{43}
\beq
Z(\omega)=-\frac{i\omega}{\omega_p}, \qquad
Z=\frac{\xi}{{\tilde{\omega}}_p}.
\label{20}
\eeq
\ni
In this case the skin depth is expressed as
$\delta=\delta_0\equiv c/\omega_p$ and thereby does not
depend on frequency.

In Ref.\cite{29} the approximate representation of the impedance
function (\ref{20}) was used to calculate perturbatively the
Casimir force between two parallel plates given by Eq.(\ref{10}).
The result was represented as a series
\beq
F_{pp}^{(0)}(a)=F_{pp}^{(0,0)}(a)
\sum\limits_{k}c_k\left(\frac{\delta_0}{a}\right)^k,
\label{21}
\eeq
\ni
where the relative skin depth $\delta_0/a$ is a small parameter
and the Casimir force between ideal metals is
$F_{pp}^{(0,0)}(a)=-\pi^2\hbar c/(240a^4)$,
in accordance with Eq.(\ref{6}). For the first coefficients
the following values were found in \cite{29}
\beq
c_0=1,\quad c_1=-\frac{16}{3},\quad c_2=24.
\label{22}
\eeq
\ni
It is notable that the same coefficients were found by using
Lifshitz theory \cite{16,40}, i.e. when instead of (\ref{9}) 
the quantities (\ref{11}) are substituted  into Eq.(\ref{10}).
In the Lifshitz formula the plasma model representation for the
dielectric permittivity which was used corresponds to the more exact
Eq.(\ref{19}) for the impedance. Expansion coefficients of higher
orders were also obtained from Lifshitz theory (up to
the fourth order in \cite{16} and up to the sixth order in \cite{44})
for both energy density and force. By way of example, in the 
framework of Lifshitz theory it follows \cite{16,44} that
\beq
c_3^{(L)}=-\frac{640}{7}\left(1-\frac{\pi^2}{210}\right)
\approx -87.13, 
\qquad
c_4^{(L)}=\frac{2800}{9}\left(1-\frac{163\pi^2}{7350}\right)
\approx 243.01,
\label{23}
\eeq
\ni
whereas by considering the impedance approach, with the more exact 
Eq.(\ref{19}), it is not difficult from Eq.(\ref{10}) to arrive at
\beq
c_3^{(i,e)}=-\frac{640}{7}\left(1+\frac{\pi^2}{280}\right)
\approx -94.65, 
\qquad
c_4^{(i,e)}=\frac{2800}{9}\left(1+\frac{5\pi^2}{294}\right)
\approx 363.33.
\label{24}
\eeq
\ni
It is seen that there are differences between the numerical
values of the higher order coefficients obtained from
the impedance approach and those obtained from Lifshitz 
theory, even if one uses the more exact expression for the impedance.
If we use the approximate expressions of Eq.(\ref{20}) the higher
order coefficients calculated in frames of the impedance approach are
\beq
c_3^{(i,a)}=-\frac{11520}{7\pi^4}\left[\zeta(3)+
\frac{1}{8}\zeta(5)\right]
\approx -22.50, 
\qquad
c_4^{(i,a)}=\frac{14000}{3\pi^4}\left[\zeta(3)+
\frac{1}{2}\zeta(5)\right]
\approx 82.43,
\label{25}
\eeq
\ni 
where $\zeta(z)$ is the Riemann zeta function.
That is in even worse agreement with the values of Eq.(\ref{23})
obtained from Lifshitz theory. The perturbation
results obtained for the energy density between the plates (i.e.
for the force between a plate and a spherical lens)
are in perfect analogy with the above ones.

Largely, however, the results obtained from the
impedance approach and those from Lifshitz theory agree
closely with each other in a wide separation range. The measure of
the agreement between both approaches is represented by the quantity
\beq
\delta F_{pp}^{(0)}(a)=
\frac{F_{pp,L}^{(0)}(a)-F_{pp}^{(0)}(a)}{F_{pp,L}^{(0)}(a)},
\label{26}
\eeq
\ni
where the Casimir force between the plates $F_{pp}^{(0)}(a)$
is calculated from Eq.(\ref{10}) (impedance approach), and
$F_{pp,L}^{(0)}(a)$ is computed in the framework of Lifshitz theory
(by the use of Eq.(\ref{11})).
The results calculated for $Al$
($\omega_p\approx 1.9\times 10^{16}\,$rad/s \cite{45}) are
presented in Fig.~1 (computations were performed by the use of
Mathematica).

Curve 1 corresponds to the more exact impedance function of Eq.(\ref{19}),
whereas curve 2 is computed by using Eq.(\ref{20}). As it is seen
from curve 1, for all separations from 100\,nm to 10\,$\mu$m
the impedance approach is in good agreement with Lifshitz theory.
The largest deviation (smaller than 0.5\%) occurs at the smallest 
separation only. Regarding the approximate impedance function of
Eq.(\ref{20}) (curve 2), the error of impedance method exceeds 5\%
at separations smaller than 200\,nm. However, for separations larger
than 1.2\,$\mu$m the relative error is less than 1\%.

For the energy density between the plates 
(force in a configuration of a spherical
lens above a plate) the impedance approach leads to even more exact results.
Here the error can be characterized by
\beq
\delta E_{pp}^{(0)}(a)=
\frac{E_{pp,L}^{(0)}(a)-E_{pp}^{(0)}(a)}{E_{pp,L}^{(0)}(a)},
\label{27}
\eeq
\ni
where $E_{pp}^{(0)}(a)$ is determined by Eq.(\ref{8}). The results
computed by the use of Mathematica 
are presented in Fig.~2 (curve 1 corresponds to the impedance
function of Eq.(\ref{19}) and curve 2 is based on the approximate
Eq.(\ref{20})). As is seen from curve 1, the results for the impedance
approach practically coincide with the results of Lifshitz
theory in the separation range 
$0.4\,\mu\mbox{m}\leq a\leq 10\,\mu$m.
In the separation range
$0.1\,\mu\mbox{m}\leq a\leq 0.4\,\mu$m 
the error introduced by the impedance approach is less than 0.3\%
(this maximal error occurs at the smallest separation).
With the approximate impedance function of Eq.(\ref{20}) the error
introduced by the impedance approach is less than 1\% at
$a\geq 0.7\,\mu$m, and does not exceed 5\% at the separations
$100\,\mbox{nm}\leq a\leq 700\,$nm.

Thus the impedance approach given by Eqs.(\ref{8}), (\ref{10}), (\ref{12}),
and (\ref{19}) seems to provide a good description of 
the Casimir force between metals
at zero temperature with required accuracy. 

\section{Casimir force in the region of
infrared optics at nonzero temperature}

In the impedance approach the Casimir energy density and force 
at nonzero temperature are given by Eqs.(\ref{15})--(\ref{17}).
If we change in these formulas $X^{\|,\bot}$ by $X_L^{\|,\bot}$,
defined by Eq.(\ref{11}), one obtains the Lifshitz results for the
Casimir energy density and force.
By applying the Poisson summation formula it is possible to
rewrite Eqs.(\ref{15}), (\ref{17}) in the form of separated 
contributions of zero temperature (as it is done in \cite{40}
for the case of the Lifshitz theory)
\beq
E_{pp}^{(T)}(a)=E_{pp}^{(0)}(a)+\Delta_TE_{pp}(a), 
\qquad
F_{pp}^{(T)}(a)=F_{pp}^{(0)}(a)+\Delta_TF_{pp}(a).
\label{28}
\eeq
\ni
Here the zero-temperature contributions are defined by Eqs.(\ref{8}),
(\ref{10}) and were computed in Sec.III.
The temperature corrections in the impedance approach are given by
\bes
&&
\Delta_TE_{pp}(a)=\frac{\hbar c}{16\pi^2a^3}
\sum\limits_{l=1}^{\infty}
\int_{0}^{\infty} y\,dy
\int_{0}^{y} d\xi
\cos\left(l\xi\frac{T_{eff}}{T}\right)
\left\{\vphantom{\ln\left[1+\frac{X^{\|}(y,\xi)}{e^y-1}\right]}
2\ln\left(1-e^{-y}\right)\right.
\nn\\
&&
\phantom{E_{pp}^{(0)}(a)}+\left.
\ln\left[1+\frac{X^{\|}(y,\xi)}{e^y-1}\right]+
\ln\left[1+\frac{X^{\bot}(y,\xi)}{e^y-1}\right]\right\},
\label{29} \\
&&
\Delta_TF_{pp}(a)=
-\frac{\hbar c}{16\pi^2a^4}
\sum\limits_{l=1}^{\infty}
\int_{0}^{\infty} y^2\,dy
\int_{0}^{y} d\xi
\cos\left(l\xi\frac{T_{eff}}{T}\right)
\nn \\
&&\phantom{aaaaaaaaa}
\times
\left[\frac{1-X^{\|}(y,\xi)}{e^y-1+X^{\|}(y,\xi)}+
\frac{1-X^{\bot}(y,\xi)}{e^y-1+X^{\bot}(y,\xi)}\right].
\label{30}
\ees

Let us start with the temperature correction to the energy density and
expand (\ref{29}) up to the second power in the small parameter $\delta_0/a$.
Note that this expansion does not depend on whether we use the more
exact Eq.(\ref{19}) or the approximate Eq.(\ref{20}) for the impedance
function. In both cases the result is
\bes
&&
\Delta_T E_{pp}(a)
=-\frac{\hbar c}{8\pi^2 a^3}
\sum\limits_{l=1}^{\infty}
\left\{
\frac{\pi}{2(lt)^3}\coth(\pi lt)-\frac{1}{(lt)^4}+
\frac{\pi^2}{2(lt)^2}\frac{1}{\sinh^2(\pi lt)}
\right.
\label{31} \\
&&\phantom{aaa}
+\frac{\delta_0}{a}\left[\frac{\pi}{(lt)^3}\coth(\pi lt)
-\frac{4}{(lt)^4}+
\frac{\pi^2}{(lt)^2}\frac{1}{\sinh^2(\pi lt)}+
\frac{2\pi^3}{lt}\frac{\coth(\pi lt)}{\sinh^2(\pi lt)}
\right]
\nn \\
&&\phantom{aaa}
-\left(\frac{\delta_0}{a}\right)^2
\left[\frac{\pi}{(lt)^5}+
\frac{2\pi^4}{\sinh^2(\pi lt)}
\left(1-3\coth^2(\pi lt)+
\frac{\coth(\pi lt)}{\pi lt}-\frac{1}{(\pi lt)^2}\right)
\right.
\nn \\
&&\phantom{aaaaa}
\left.\left.
+\frac{6}{\pi(lt)^5}\left(
2\pi lt\ln\left(1-e^{-2\pi lt}\right)-
\frac{2\pi^2 (lt)^2}{e^{2\pi lt}-1}-
{\mbox{Li}}_2\left(e^{-2\pi lt}\right)\right)
\right]
\vphantom{\frac{\pi^3}{\sinh^2(\pi lt)}}
\right\},
\nn
\ees
\ni
where Li$(z)$ is the polylogarithmic function  and 
$t\equiv T_{eff}/T$.
It is remarkable that Eq.(\ref{31}) coincides with the temperature
correction obtained from Lifshitz theory in the
same approximation \cite{41}. Because of this, the impedance
approach, when applied to calculate the temperature correction
to the Casimir force, produces even more exact results than at zero
temperature. In fact, the results computed by 
the perturbative Eq.(\ref{31})
coincide with those computed numerically by the exact Lifshitz
formula up to four significant figures in all separation range
under consideration, from 0.1\,$\mu$m to 10\,$\mu$m.

Now we consider the temperature correction to the Casimir force
given by Eq.(\ref{30}). Expanding (\ref{30}) in
a power series of the small parameter $\delta_0/a$ and keeping
terms up to the second order one obtains
\bes
&&
\Delta_T F_{pp}(a)
=-\frac{\hbar c}{8\pi^2 a^4}
\sum\limits_{l=1}^{\infty}
\left\{\frac{1}{(lt)^4}-
\frac{\pi^3}{lt}\frac{\coth(\pi lt)}{\sinh^2(\pi lt)}
\right.
\nn \\
&&\phantom{aaa}
+\frac{\delta_0}{a}\frac{\pi^3}{lt\sinh^2(\pi lt)}
\left[\frac{1}{(\pi lt)^2}\sinh(\pi lt)\cosh(\pi lt)+
4\coth(\pi lt)
\right.
\nn \\
&&\phantom{aaaaaaaaaaaaaa}
\left.
+2\pi lt-
6\pi lt\coth^2(\pi lt)+\frac{1}{\pi lt}\
\right]
\label{32} \\
&&\phantom{aaa}
+3\left(\frac{\delta_0}{a}\right)^2\frac{\pi^3}{lt\sinh^2(\pi lt)}
\left[-4\pi lt +5(\pi lt)^2\coth(\pi lt)+12\pi lt\coth^2(\pi lt)
\right.
\nn \\
&&\phantom{aaaaaaaaaa}
\left.\left.
-8(\pi lt)^2 \coth^3(\pi lt)
-4\coth(\pi lt)\right]
\vphantom{\frac{\pi^3}{\sinh^2(\pi lt)}}
\right\}.
\nn
\ees
\ni
This result is obtained for the representation (\ref{19}) as well 
as for representation (\ref{20}) of the impedance function.
It coincides with the perturbative calculation of the temperature
correction to the Casimir force in Lifshitz theory.
Once more, the results of numerical computations based on
Lifshitz theory agree with (\ref{32}) up to four sifnificant
figures in the separation range from 0.1\,$\mu$m to 10\,$\mu$m.
This means that the  impedance approach is well suited for 
calculating the temperature correction to the Casimir force, leading to 
practically exact results in a wide separation range. 

\section{Casimir force in the region of
the normal skin effect}

In this Section, the case of large separations between the test bodies
$a>1\,$mm is considered. For such separations the characteristic
frequencies giving the crucial contribution to the Casimir force 
correspond to the normal skin effect. Although at so large separations 
the Casimir force is very small and at present cannot be detected
experimentally, some special features make this case interesting
from a theoretical point of view.

In the region of the normal skin effect the surface impedance is complex,
namely
\beq
Z(\omega)=(1-i)\sqrt{\frac{\omega}{8\pi\sigma}},
\label{33}
\eeq
\ni
where $\sigma$ is the conductivity of the boundary metal. It can be related
to the effective plasma frequency by $\sigma=\omega_p^2/(4\pi\gamma)$
($\gamma$ is the relaxation parameter). Note that in the region of the
normal skin effect the characteristic frequencies of the Casimir effect are
much smaller than the relaxation parameter. 
Because of this the dissipation processes play an important role in the
region of the normal skin effect and cannot be neglected.
In terms of dimensionless
quantities the impedance function (\ref{33}) on the imaginary axis can
be represented as
\beq
Z\equiv Z\left(i\frac{c\xi}{2a}\right)=
\sqrt{\frac{\xi}{4\pi{\tilde{\sigma}}}},
\label{34}
\eeq
\ni
where $\tilde{\sigma}=2a\sigma/c$. The natural small parameter of the
problem is
\beq
\sqrt{\frac{1}{4\pi{\tilde{\sigma}}}}=
\frac{1}{\sqrt{8\pi}}\sqrt{\frac{c}{\sigma a}}=
\frac{1}{\sqrt{2}}\sqrt{\frac{\delta_0}{a}\frac{\gamma}{\omega_p}}\ll 1,
\label{35}
\eeq
\ni
where $\delta_0$ is the penetration depth in the domain of 
infrared optics (see Sec.III). Taking into account that $\delta_0$
is of the order of several tens of nanometers and $\gamma$ is usually two
or three orders smaller than $\omega_p$, one can conclude that the parameter
(\ref{35}) is smaller than $10^{-3}$.

Consider first the case of zero temperature. Substituting (\ref{34}) into
Eqs.(\ref{8})--(\ref{10}) and expanding up to the first order in the small
parameter (\ref{35}) one obtains
\beq
E_{pp}^{(0)}(a)=E_{pp}^{(0,0)}(a)
\left[1-\frac{405\sqrt{2}}{4\pi^4}\,\zeta\left(\frac{7}{2}\right)
\sqrt{\frac{c}{\sigma a}}\right],
\qquad
F_{pp}^{(0)}(a)=F_{pp}^{(0,0)}(a)
\left[1-\frac{945\sqrt{2}}{8\pi^4}\,\zeta\left(\frac{7}{2}\right)
\sqrt{\frac{c}{\sigma a}}\right].
\label{36}
\eeq
\ni
Note that the
coefficients near the expansion parameter in (\ref{36}) are
approximately equal to 1.656 and 1.932, respectively.
The second of them (for the force) was first computed
numerically in \cite{29} (see also \cite{20}).
As an example, for $Al$ 
($\gamma\approx 9.6\times 10^{13}\,$rad/s \cite{45}) 
at $a=1\,$mm the correction
to unity in Eq.(\ref{36}) is approximately equal to 
$1.6\times 10^{-3}$ (for the energy density) and 
$1.9\times 10^{-3}$ (for the force), i.e., much smaller than 1\%.
Precisely the same results are obtained by the numerical computations 
considering the Lifshitz formula and the Drude model representation
for the dielectric permittivity. Thus at zero temperature the
impedance approach and the Lifshitz formula applied in the region 
of the normal skin effect are in agreement. 

Now consider the case of nonzero temperature. This case as described 
by Lifshitz theory presents difficulties because in the region
of the normal skin effect dissipation processes play an important
role. According to the results of Refs.\cite{12,41} in the presence of
dissipation the zeroth term of the Lifshitz formula for the Casimir force
at nonzero temperature is not well-defined. The direct application
of this formula leads to non-physical results. In fact, as shown
in Refs.\cite{2,12,41}, the rigorous derivation of the Lifshitz
formula in the framework of the Temperature Quantum Field Theory in 
Matsubara formulation leaves indefinite the zero-frequency contribution
to the Casimir force in dissipative media. Thus a special prescription
is needed to describe the Casimir effect at nonzero temperature in the
region of the normal skin effect basing on the Lifshitz formula.

By contrast, surface impedance approach can be immediately applied
in the case of nonzero temperature and does not need any prescription.
It follows from Eqs.(\ref{9}) and (\ref{34}) that
\beq
X^{\|}(y,0)=X^{\bot}(y,0)=0,
\label{37}
\eeq
\ni
and the zero frequency contribution to the energy density (\ref{15})
and force (\ref{17}) are the same as for ideal metal. The other terms in
Eqs.(\ref{15}, (\ref{17}) also can be calculated without trouble.
The results for $Al$ at $a=1\,$mm are as follows.

For large separations which are typical for the normal skin effect 
the effective temperature defined in Eq.(\ref{14}) is very low 
(to illustrate, $T_{eff}=1.145\,$K at $a=1\,$mm). Since it is not
reasonable to further increase the separation distance we consider
the dependence of the Casimir energy density and force on
temperature at a fixed separation. The results are found to be very
close to the ones obtained for an ideal metal. At $T=1\,$K
the value of $E_{pp}^{(T)}$ computed by the impedance method is
equal to 0.99992 of the value computed by Eq.(\ref{16}) for an 
ideal metal (to compare, the same quantity computed from the
Lifshitz formula appended by the special prescription \cite{12}
is only 0.005\% lower). For $T\geq 2\,$K the impedance approach
leads to practically the same values as obtained for an ideal metal.
The Casimir force $F_{pp}^{(T)}$ at $T=1\,$K computed by the
impedance approach is equal to 0.999745 of the force between
ideal metals. Results of this kind are quite natural in the
range of the normal skin effect because at so large separations any
metal behaves like an ideal one.

\section{Conclusion and discussion}

As indicated above, the method of the surface impedance gives
the possibility of calculating the Casimir energy density and force
at both zero and nonzero temperature. The important advantage
of this method lies in its simplicity. As distinct from Lifshitz
theory the realistic properties of the test body material are 
included into the boundary condition and the electromagnetic
fluctuations inside the media are not considered. Surface impedance
is a more general characteristic than the dielectric permittivity.
As a result, the impedance approach may be employed in  cases
in which the Casimir force cannot be calculated from
 Lifshitz theory.

In the paper the main concepts of the impedance approach
to the theory of the Casimir effect were outlined. The formal 
substitution which enables one to link the formalism of
the impedance approach with the Lifshitz theory was found. 
The expressions
for the Casimir energy density and force at nonzero temperature
were obtained first in terms of the surface impedance (Sec.II).
The obtained expressions were applied to calculate the Casimir force 
in the configuration of two parallel plates and a sphere (spherical 
lens) above a plate in two important cases. The first case corresponds
to the space separations between the test bodies of order of one
micrometer. Here the visible and infrared frequencies contribute
most to the Casimir force. This case is important from an
experimental point of view. The second case corresponds to 
separations larger than one millimeter, which is the region of 
the normal skin effect.

The results obtained in the impedance approach were
carefully compared with those calculated using
Lifshitz theory. In the separation range
$0.1\,\mu\mbox{m}\leq a\leq 10\,\mu$m at zero temperature the
deviation between both approaches is less than 0.5\% (for two
parallel plates) and 0.3\% (a sphere above a plate) when the 
impedance function of Eq.(\ref{19}) is used. The maximal errors
mentioned above occur at the smallest separation.
At $a\geq 400\,$nm the results are practically coincident with
those of Lifshitz theory.

The impedance approach was shown to be well suited for calculating
the temperature correction to the Casimir force. In the region of the
visible light and infrared optics it leads to the same perturbative 
results as the Lifshitz formula (see Sec.IV). The agreement between
the impedance approach and the numerical computations in
Lifshitz theory, at least up to four significant figures,
was demonstrated.

Although at separations larger than 1\,mm (the region of the normal
skin effect) the Casimir force is too small to be observed 
experimentally, this case is interesting from a methodological point 
of view. At zero temperature the corrections to the Casimir 
energy density and force were found analytically (Eq.(\ref{36})).
They are in agreement with the numerical computations performed
by using the Lifshitz formula. At nonzero temperature the case of the
normal skin effect presents difficulties when Lifshitz theory
is used to describe the Casimir force. The reason is that in the 
presence of dissipation the zero-frequency term of the Lifshitz
formula is not well-defined. As shown in Sec.V, the impedance
approach does not present any difficulty when applied to calculate
the Casimir force in the region of the normal skin effect.
In Sec.V practically the same results for the Casimir force
were obtained in the impedance approach as for an ideal metal.
These results are approximately equal to those obtained from the
Lifshitz formula appended by special prescription.

In conclusion, it may be said that the impedance approach to the
theory of the Casimir effect is rather effective and provides
the means for precise calculations of the Casimir force
acting between realistic media in a wide separation range.
 
\section*{Acknowledgments}

    The authors are grateful to U.~Mohideen and V.M.~Mos\-te\-pa\-nen\-ko
for helpful discussions. 
G.L.K. is indebted to the Department of Physics
of the Federal University of Paraiba, where this work 
was done, for kind hos\-pi\-ta\-li\-ty. 
Financial support from CNPq is acknowledged.

\normalsize
\newpage
\begin{figure}[h]
\vspace*{-7cm}
\epsfxsize=20cm\centerline{\epsffile{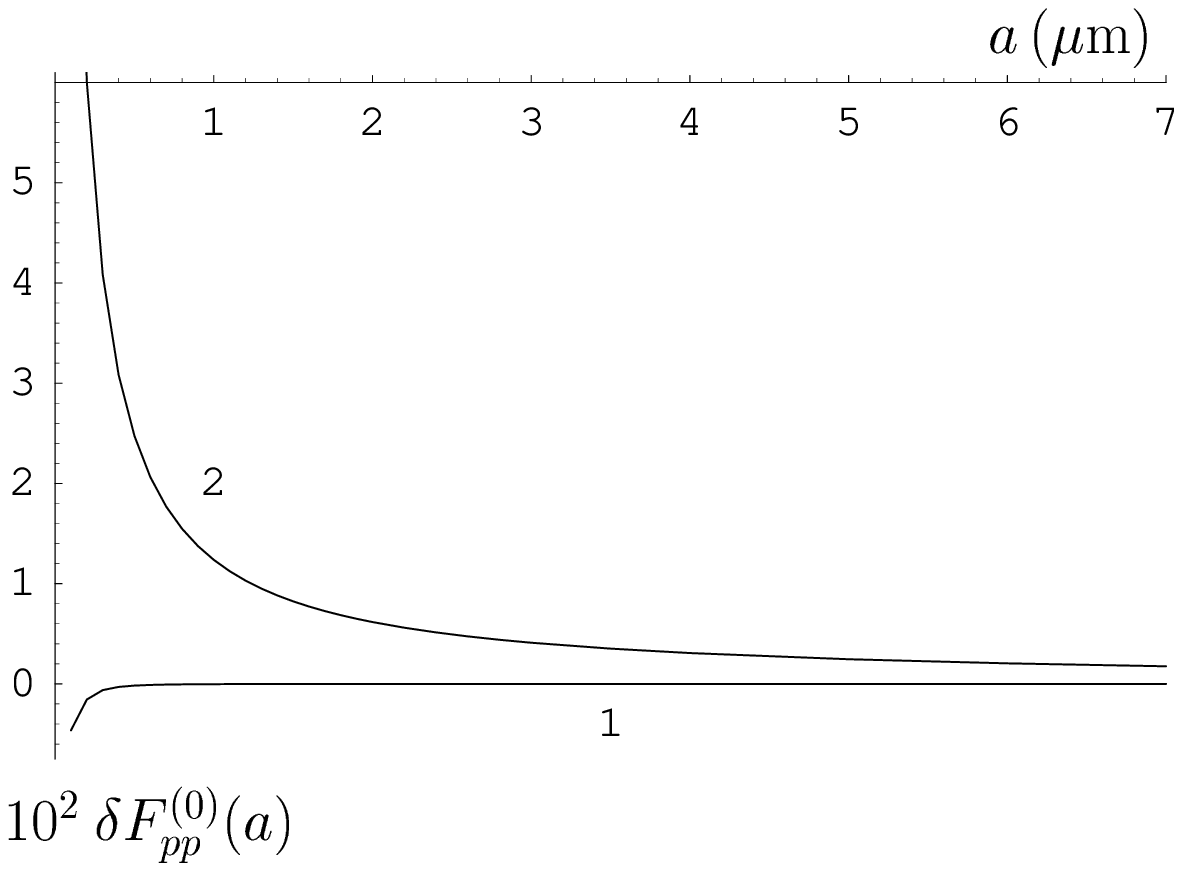}}
\vspace*{-8cm}
\caption{The relative deviation of the Casimir force
between parallel plates at zero temperature as computed
in the impedance approach and from 
Lifshitz theory. Curve 1 corresponds to the impedance
function of Eq.(19) and curve 2 to the impedance of Eq.(20).
}
\end{figure}
\newpage
\begin{figure}[h]
\vspace*{-7cm}
\epsfxsize=20cm\centerline{\epsffile{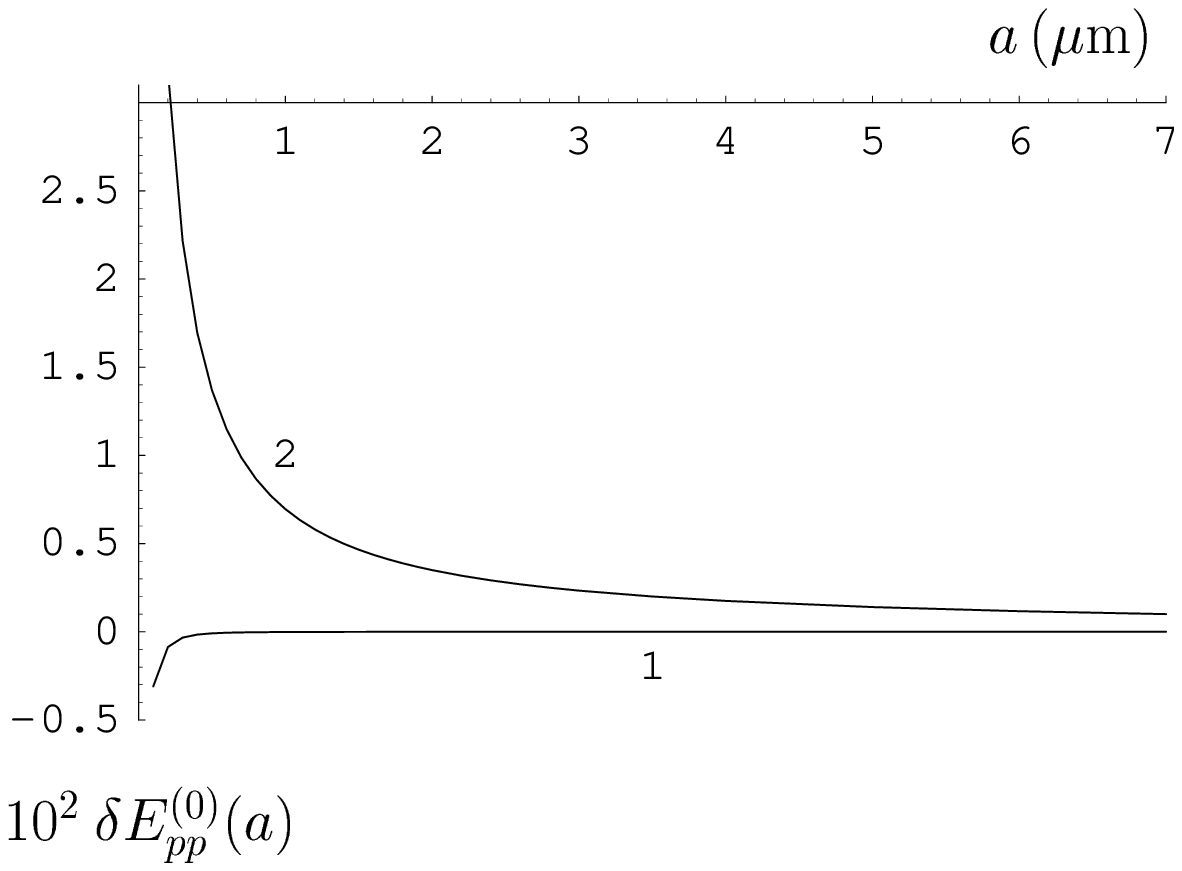}}
\vspace*{-8cm}
\caption{The relative deviation of the Casimir energy density
between parallel plates at zero temperature as computed
in the impedance approach and from
Lifshitz theory. Curve 1 corresponds to the impedance
function of Eq.(19) and curve 2 to the impedance of Eq.(20).
}
\end{figure}

\end{document}